# Recent advances in computational methods for studying ligand binding kinetics

**Authors:** Farzin Sohraby, Ariane Nunes-Alves

Institute of Chemistry, Technische Universität Berlin, 10623 Berlin, Germany

*Correspondence: ferreira.nunes.alves@tu-berlin.de (Nunes-Alves, A.)

## Abstract

Binding kinetic parameters can be correlated with drug efficacy, which led to the development of various computational methods for predicting binding kinetic rates and gaining insight into protein-drug binding paths and mechanisms in recent years. In this review, we introduce and compare computational methods recently developed and applied to two systems, trypsin-benzamidine and kinase-inhibitor complexes. Methods involving enhanced sampling in molecular dynamics simulations or machine learning can be used not only to predict kinetic rates, but also to reveal factors modulating the duration of residence times, selectivity and drug resistance to mutations. Methods which require less computational time to make predictions are highlighted, and suggestions to reduce the error of computed kinetic rates are presented.

## Binding kinetics as an indicator of drug efficacy in vivo

In recent years, it became apparent that calculating binding kinetic rates and understanding the binding mechanism of drugs to their target proteins can be helpful in drug design [1–3]. For instance, recent works showed that the residence time (Box 1) can, in many cases, be better correlated with drug efficacy in vivo, compared to equilibrium descriptors [4,5]. This led to a growing interest in the development of computational methods for gaining insight into binding kinetics and mechanisms, and to this date many methods have been developed for this purpose.

In this review, we briefly introduce and compare methods published in recent years, from 2019 onward. We mainly focus on two protein types: trypsin and kinases (Abelson (Abl), Src, p38 mitogen activated protein (MAP), and focal adhesion kinase (FAK)). These proteins were chosen due to the availability of many publications applying computational methods to study protein-ligand association and dissociation for them, which facilitates comparison across different methods. The trypsin-benzamidine complex was used to benchmark new methods, especially methods which enhance sampling in molecular dynamics (MD) simulations. On the other hand, large data sets, with tens or hundreds of inhibitors with experimentally characterized kinetic rates, are available for kinases, allowing the development and application not only of enhanced sampling methods, but also of machine learning or chemometric approaches. The results obtained using these proteins are compared and summarized below.

In this brief review, it is not possible to describe all relevant methods and publications. We refer the reader to the on-line toolbox Kbbox [6], which contains descriptions of different methods, and to recent reviews [7–14].

## The trypsin-benzamidine complex as a model system to benchmark new computational methods

Trypsin in complex with benzamidine (Figure 1) is a model system that has been extensively used for developing different enhanced sampling methods to bridge the difference in time scales between trypsin-benzamidine dissociation (experimental residence time of 1.7 ms [15]) and MD simulations (usually limited to tens of microseconds). Many of the recently developed methods are modifications of previous methods, or combinations of different approaches such as weighted ensemble (WE) [16] and milestoning [17]. Such methods will be referred to as second-generation methods.

## Accurate kinetic rates can be obtained with less than one microsecond of simulation time

A total of eleven studies applied enhanced sampling methods to study ligand binding to trypsin [18–28]. There is no clear correlation between simulation time and error of predicted $k_{off}$ values (Figure 2, Table 1). Here we define as accurate a method which has an error (fraction of the computed rate over the experimental rate) of 0.5 to 1.5-fold. Notably, two second-generation methods were able to predict accurate absolute $k_{off}$ values with less than one microsecond of total simulation time, Markovian Weighted Ensemble Milestoning (M-WEM) [20] and dissipation-corrected Targeted Molecular Dynamics (dcTMD) [22]. M-WEM is based on weighted ensemble milestoning (WEM) [29], while dcTMD combined with Langevin simulations [22] is based on TMD [30].

## Second-generation methods can lead to predictions of kinetic rates with lower errors and improved sampling

The methods Simulation Enabled Estimation of Kinetic Rates 2 (SEEKR2) [18] and Markovian Milestoning with Voronoi Tessellations (MMVT) SEEKR [19], based on SEEKR [31], predicted kinetic rates for the trypsin-benzamidine complex with low error (Figure 2, Table 1). Notably, SEEKR2 predicted kinetic rates with lower error than its predecessor, MMVT SEEKR. SEEKR uses MD simulation for milestones close to the binding pocket, where atomistic details matter, and Brownian Dynamics (BD) simulation for milestones further from the binding pocket to reduce the amount of the simulation time to compute kinetic rates. In MMVT SEEKR [19] the MMVT approach was introduced to overcome the need to sample equilibrium distributions for each milestone, leading to a drastic reduction in computational time. After MMVT SEEKR, a newer version of this method was developed, SEEKR2 [18], which includes more options of software to perform MD simulations (openMM, in addition to NAMD), leading to a gain in computational performance for MD simulations from 47 ns/day to 300 ns/day. Nonetheless, both MMVT SEEKR and SEEKR2 require a few microseconds of simulation to make predictions. SEEKR2 estimated a $k_{on}$ value with lower error than MMVT SEEKR (Table 1), which can be explained by small differences in the implementation of the methods. For instance, the milestones in SEEKR2 were adjusted, and desolvation forces were included in the BD simulations.

Resampling of Ensembles by Variation Optimization (REVO) [23] is a new method to enhance sampling based on WExplore [32]. REVO and WExplore are based on WE, which uses an ensemble of trajectories to sample (un)binding events. REVO contains a modified algorithm to maximize variation in the ensemble of trajectories. In addition to sampling two unbinding paths identified by WExplore, REVO found two more paths (Figure 1), which indicates that REVO does a better exploration of the conformational space.

Interestingly, the $k_{off}$ values estimated by WExplore and REVO are close to the experimental one (Table 1), despite the fact that WExplore sampled fewer unbinding paths. This indicates that exhaustive pathway sampling is not required to predict kinetic rates with low error, as long as the main pathway is sampled in the simulations, in line with what was recently observed for T4 lysozyme [33].

**Force field choices and insufficient sampling can lead to high errors in the predicted rates**

Another second-generation method is the Variational Approach to Conformational dynamics Metadynamics (VAC-MetaD) combined with infrequent metadynamics (InMetaD) to compute kinetic rates [24]. VAC-MetaD was used to choose and optimize collective variables, while InMetaD was used to increase sampling of dissociation events. The $k_{off}$ value resulting from this method is about 7 times higher than the experimental value (Table 1). The authors attributed this high $k_{off}$ value to the parameters used for benzamidine, which facilitated rupture of trypsin-benzamidine interactions in the bound state, therefore speeding up dissociation. This is an example of how force fields can affect the accuracy of computed kinetic rates, as highlighted in previous studies [11,34].

One of the challenges in obtaining protein-ligand association and dissociation events is the time scales achieved by conventional MD simulations. One of the ways to extend such time scales is using coarse-grained MD (CGMD) simulations. Dandekar et al. [25] captured trypsin-benzamidine binding pathways using CGMD simulations with an aggregate simulation time of 428 microseconds (Table 1). The predicted kinetic rates are far from the experimental values, compared to the predictions with all-atom force fields (Table 1). The high errors in the predictions can be explained by the facilitated binding and unbinding events in CGMD simulations due to a smoothed potential energy, as pointed out in a previous study [35]. Such smoothing can facilitate conformational changes and reduce barriers for un(binding). In order to compute accurate kinetics rates from CGMD simulations, a scaling factor may be needed to correct for the smoothed potential energy.

Ligand Gaussian accelerated Molecular Dynamics (LiGaMD) [21], which is based on Gaussian accelerated Molecular Dynamics (GaMD) [36], is a new enhanced sampling method which selectively boosts the nonbonded potential energy of the ligand to achieve dissociation events. A dual-boost algorithm can be applied, where the total potential energy of the system is also boosted to facilitate the occurrence of binding events. Moreover, multiple ligands are included in the simulation box to increase the likelihood of binding events, similar to what is done in the method Unaggregated Unbiased MD (UUMD) [37], which includes additionally a small repulsive force between virtual interaction sites on the ligands to avoid aggregation. Using LiGaMD and the dual-boost scheme, the $k_{on}$ value obtained was on the same order of magnitude of the experimental $k_{on}$, but the estimated $k_{off}$ value had a higher error than other methods (Table 1, Figure 2). This can be an indication that improved sampling of the bound state is required to obtain better predictions of $k_{off}$ values.

**Table 1.** Errors and simulation times of different methods for computing association ($k_{on}$) and dissociation rates ($k_{off}$) for the trypsin-benzamidine complex and for kinase-inhibitor complexes.

| Protein | Ligand | Method | $k_{off}$ ($s^{-1}$) | $k_{on}$ ($10^7$ $M^{-1}$ $s^{-1}$) | Error (fold)[a] - $k_{off}$ | Error (fold)[a] - $k_{on}$ | simulation time (µs)[b] | ref. |
|---|---|---|---|---|---|---|---|---|
| **Trypsin** | **Benzamidine** | **Experiment** | **600** | **2.90** | **1.00** | **1.00** | **-** | [15] |
| Trypsin | Benza | M-WEM | 791 ± | 0.53 ± | 1.32 | 0.18 | 0.48 | [20] |
| | midine | | 197 | 0.08 | | | | |
| Trypsin | Benza midine | dcTMD | 270 ± 40 | 0.87 ± 0.05 | 0.45 | 0.30 | 0.80 | [22] |
| Trypsin | Benza midine | InMetaD and VAC-MetaD | 4176 ± 324 | - | 6.96 | - | ~ 1.00 | [24] |
| Trypsin | Benza midine | MMVT SEEKR | 174 ± 9 | 12 ± 0.5 | 0.29 | 4.14 | 4.40 | [19] |
| Trypsin | Benza midine | SEEKR2 | 990 ± 130 | 2.4 ± 0.2 | 1.65 | 0.83 | 5.00 | [18] |
| Trypsin | Benza midine | LiGaMD | 3.53 ± 1.41 | 1.15 ± 0.79 | 0.006 | 0.40 | 5.00 | [21] |
| Trypsin | Benza midine | WExplor e | 840 | - | 1.40 | - | 8.75 | [23] |
| Trypsin | Benza midine | REVO | 266 | - | 0.44 | - | 8.75 | [23] |
| Trypsin | Benza midine | CGMD | $6.9\cdot10^{5}$ | 36.8 | 1150 | 12.69 | 428 | [25] |
| **Src kinase** | **Dasati nib** | **Experim ent** | **0.06** | **0.50** | **1.00** | **1.00** | **-** | [38] |
| Src kinase | Dasatin ib | UUMD | - | 0.76 | - | 1.52 | 6.6 | [37] |
| Src kinase | Dasatin ib | SuMD | $5.3\cdot10^{5}$ | - | $8.8\cdot10^{6}$ | - | 4.4 | [39] |
| Src kinase | Dasatin ib | CGMD | - | 4.00 | - | 8.00 | 300 | [40] |
| **Abl kinase** | **Imatini b** | **Experim ent** | **8.3 ± $0.8\cdot10^{-4}$ / 25 ± 6** | **-** | **1.00 / 1.00** | **-** | **-** | [41] / [42] |
| **Abl kinase (N368S)** | **Imatini b** | **Experim ent** | **2.7 ± $0.3\cdot10^{-3}$** | **-** | **1.00** | **-** | **-** | [41] |
| Abl kinase | Imatinib | Milestoni ng | 18 | - | $2.2\cdot10^{4}$ / 0.73[c] | - | 0.82 | [43] |
| Abl kinase | Imatinib | InMetaD | 6 ± $3\cdot10^{-4}$ | - | 0.72 / $2.4\cdot10^{-5}$[c] | - | NA[d] | [44] |
| Abl kinase (N368S) | Imatinib | InMetaD | 4 ± $2\cdot10^{-3}$ | - | 1.48 | - | NA[d] | [44] |

[a]Fraction computed/predicted rate over experimental rate. [b]Only production MD simulations are considered. System preparation and equilibration are excluded from this calculation. [c]Error values calculated based on both experimental $k_{off}$ values available for the Abl-Imatinib complex. [d]Not available.

**Large data sets for kinase-inhibitor complexes as model systems to train and test methods**

Protein kinases (Figure 3) are involved in nearly all regulatory pathways in cells and targeting them has been proven to be a successful strategy to fight many diseases [45,46]. Due to their importance and the availability of many experimental data, such as kinetic rates and crystal structures, proteins such as Abl kinase, Src kinase, FAK and p38 MAP kinase are among the most studied kinases. In drug design, understanding the mechanistic details of their inhibition is of great interest to achieve selectivity and to address drug resistance in mutants, which led numerous computational studies to focus on these kinases. In this section, we will present the main results from recent studies concerning the above-mentioned kinases.

**Machine learning models can achieve high predictive power at low computational cost**

The machine learning models Volsurf [47] and ensemble COMparative BINding Energy (COMBINE) [48], which used one type of protein in the data set, achieved the highest predictive powers, as measured by the coefficient of determination ($R^2$) of the test set (Table 2). Volsurf is a chemometric method which makes use of partial least squares (PLS) regression to predict kinetic rates, using as features properties of the inhibitors only, which limits the mechanistic insights that can be obtained. 28 p38 MAP kinase inhibitors were considered, and the best model presented $R^2$ of 0.82 for the test set (Table 2) [47]. The COMBINE analysis method was recently extended to ensemble COMBINE [48], which allows the use of multiple structures to represent one protein-ligand complex in order to represent flexibility. The best model from ensemble COMBINE, obtained using 33 complexes with p38 MAP kinase, presented $R^2$ of 0.79 for the test set (Table 2).

Another study also applied PLS to p38 MAP kinases, obtaining a lower predictive power. Zhang et al. [49] built a model to predict $k_{off}$ values using a data set of 20 complexes (similar to the data set used in ensemble COMBINE) and protein-ligand interaction energies from dissociation trajectories obtained from position-restrained MD simulations as features. This model, which considers intermediate states along the dissociation pathways, was expected to have a higher predictive power than ensemble COMBINE, which does not consider such states. Surprisingly, the model had a lower predictive power, since it presented $R^2$ of 0.56 for the test set (Table 2). As pointed out previously [48], the fact that models which use static structures can achieve high predictive power may be explained by structural similarity between the bound state and the transition state ensemble for dissociation. Such similarity was observed before for Imatinib dissociation from Abl kinase using milestoning [43].

The availability of large data sets containing experimental kinetic rates [50–52] enabled the use of machine learning for data sets including different proteins types. Two studies [50,52] built models using random forest, data sets with 501 or 680 protein-ligand complexes and protein atom-ligand atom pairs as features. The models predicted $k_{off}$ values for their training sets with average or high accuracy ($R^2$ of 0.60 or 0.94, Table 2). However, the predictive power for the test sets (28 complexes with p38 MAP kinases or 33 complexes with FAK) was low ($R^2$ of 0.43 and 0.06, Table 2). One of the issues pointed out by the authors to explain the low predictive power was the use of static structures to predict $k_{off}$ values [52]. However, this does not seem to be a problem for models trained using one single protein type (see above). Other possible issues could be the heterogeneity of conditions and methods used to obtain the experimental data, or the low transferability of the models across different protein types or ligands with diverse structures.

Machine learning methods can be reasonably accurate in the prediction of kinetic rates and are very efficient since, after training, they can predict kinetic rates in seconds, while simulation-based methods usually take days or more to compute kinetic rates. However, methods based on machine learning usually do not reveal information about binding paths and metastable states, and require a lot of experimental data for training.

**More simulation time can lead to improved predictions for kinetic rates**

Different publications also studied binding kinetics in p38 MAP kinases using conventional MD simulations or enhanced sampling methods [53–57]. Steered MD (SMD) simulations [56] displayed a higher predictive power than local-scaled MD [53] (Table 2), at the expenses of employing more simulation time. With the goal of enhancing the sampling of ligand dissociation events in MD simulations, Du et al. applied a method called local-scaled MD , where a smooth and gradual reduction of the interaction energies between protein and ligand facilitate the occurrence of unbinding events. Local-scaled MD changes the potential energy of the system, similar to other enhanced sampling methods, such as scaled MD [58,59] and selectively scaled MD [60]. Local-scaled MD was employed to study dissociation pathways and kinetic rates in a data set of 41 diverse kinase-inhibitor complexes, resulting in $R^2$ of 0.64 for a comparison between experimental $k_{off}$ values and the binding energy integrals computed from the dissociation trajectories (Table 2). Braka et al. [56] simulated the unbinding pathways of a data set containing 8 complexes of p38 MAP kinases with inhibitors using SMD simulations. Then, by using the transition state theory and potentials of mean force (PMF), they predicted $k_{off}$ values with $R^2$ of 0.88 (Table 2), the highest $R^2$ value among the methods tested for p38 MAP kinase. The higher accuracy of this method, compared to local-scaled MD, could be due to the smaller data set used, or due to the use of a PMF to make predictions, while in local-scaled MD a more approximate computation of binding energies using molecular mechanics combined with the generalized Born surface area continuum solvation (MM/GBSA) was performed. The use of a PMF explains the higher amount of simulation time, 4.5 μs, required by SMD to make predictions. However, such method is not practical for large data sets like the one used in local-scaled MD.

**Enhanced sampling simulations provide mechanistic insights about selectivity and drug resistance**

Enhanced sampling methods were also applied to obtain mechanistic insights about selectivity and drug resistance in kinases. Berger et al. [61] used a combination of experiments and simulations to reveal the mechanism of selectivity of FAK inhibitors over the proline-rich tyrosine kinase 2 (PYK2). The enhanced sampling method tau-Random Acceleration MD (tauRAMD) computed residence times for 12 inhibitors in complex with FAK and PYK2 which reproduced the experimental ones ($R^2$ of 0.91, Table 2). The authors found that the slow dissociation of some FAK inhibitors compared to PYK2 is due to the presence of a helical conformation of the DFG motif in complexes with these inhibitors, resulting in more hydrophobic interactions and a more stable protein-ligand complex, whereas in PYK2 this conformation of the DFG motif is absent, resulting in faster dissociation for inhibitors. While previous studies [33,61–63] computed relative $k_{off}$ values using tauRAMD, a recent study also proposed a method to estimate absolute $k_{off}$ values from tauRAMD using multiple dissociation forces [64].

In an effort to obtain mechanistic insights for the resistance of the mutant N368S of Abl kinase to Imatinib, Shekhar et al. [44] performed InMetaD on both wild-type (WT) and mutant forms of Abl kinase, using AMINO and SGOOP to optimize collective variables. Using different collective variables, the authors found different dissociation pathways for each form, which they attributed to the mutation. In the WT form, Imatinib leaves the binding pocket from the hinge region, whereas in the mutant form Imatinib dissociates from the A-loop region (Figure 3B), which explains the higher $k_{off}$ value of the mutant form.

Regarding pathways for ligand dissociation, local-scaled MD [53] provided the most extensive sampling for p38 MAP kinases, with 2 paths characterized for type I inhibitors and 3 paths characterized for type II inhibitors (Figure 3A). In the study of Braka et al. [56] 1 path was observed for type I inhibitors and 2 paths were observed for type II inhibitors. Taken together, these studies suggest that there may be a relationship between long residence times, typical of type II inhibitors, and more pathways available for dissociation. On the other hand, a recent study using metadynamics and different collective variables [57] found the opposite trend for the dissociation of two type II inhibitors from p38 MAP kinase: 3 paths were found for the inhibitor with short residence time, while 1 path was found for the inhibitor with long residence time.

**The choice of method, force field or experimental reference can lead to high errors in the predicted rates**

Different studies explored the dissociation of the drug Imatinib or Gleevec from Abl kinase [43,44,65]. Narayan et al. [43] computed a $k_{off}$ value of 18 $s^{-1}$ using milestoning, while Shekhar et al. [44] obtained a $k_{off}$ value of 6 ± 3 $10^{-4}$ $s^{-1}$ using InMetaD (Table 1). The $k_{off}$ value from InMetaD is orders of magnitude different from the $k_{off}$ value estimated using milestoning. It should be noted that the studies from Shekhar et al and Narayan et al used different experimental studies to benchmark their simulations, and such studies estimated different $k_{off}$ values for the dissociation of Imatinib from Abl kinase (8.3 ± 0.8 $10^{-4}$ $s^{-1}$ in ref. [41] and 25 ± 6 $s^{-1}$ in ref. [42]). Therefore, each $k_{off}$ value predicted by the simulations can be considered to have high or low error, depending on the experimental study adopted as a reference. While the discrepancy between the experimental studies can be partially attributed to different methods and conditions, more computational and experimental studies could provide further mechanistic insights to understand such difference.

Different methods were employed to investigate the binding of drugs to Src kinase [37,39,40,66,67]. Using CGMD, Souza et al. [40] computed a $k_{on}$ value of 40 $\mu M^{-1}$ $s^{-1}$ for binding of the drug Dasatinib, which is higher than the experimental value of 5 $\mu M^{-1}$ $s^{-1}$ (Table 1), following the trend observed for CGMD applied to the trypsin-benzamidine complex (Table 1). According to the authors, this deviation can be linked to a temporal speed-up due to a smoother energy landscape of the CG force field, as discussed above.

Supervised MD (SuMD) was used to obtain unbinding of Dasatinib from Src kinase [39]. The $k_{off}$ value obtained from SuMD is orders of magnitude higher than the experimental one (Table 1). In this method, snapshots closer to unbinding events are selected for propagation, which could lead to sampling of the fastest possible dissociation events and, therefore, high $k_{off}$ values.

**Table 2**. Coefficients of determination ($R^2$), simulation times and number of complexes used to train or test different methods for computing dissociation rates ($k_{off}$) for kinase-inhibitor complexes.

| Model System | Method | Experimental $k_{off}$ ($s^{-1}$) | Number of complexes | $R^2$ training set | $R^2$ test set | Simulation time (μs)[a] | ref. |
|---|---|---|---|---|---|---|---|
| many proteins | Random Forest | $1.0\cdot10^{-6}$ - $1.0\cdot10^{2}$ | 501[b] (28)[c] | 0.60[d] | 0.43[e] | - | [50] |
| many proteins | Random Forest | $1.0\cdot10^{-7}$ - $1.0\cdot10^{3}$ | 680 (33) | 0.94[d] | 0.06[f] | $2.0\cdot10^{-3}$ - $4.0\cdot10^{-3}$ | [52] |
| many | Local- | $2.3\cdot10^{-6}$ - 4.9 | 41 | - | 0.64 | 0.25 | [53] |
| kinases | scaled MD | | | | | | |
| p38 MAP kinase | Volsurf | $1.0{\cdot}10^{-6}$ - $1.0{\cdot}10^{-1}$ | 28 (10) | 0.82 | 0.82 | - | [47] |
| p38 MAP kinase | Ensembl e COMBIN E | $8.3{\cdot}10^{-6}$ - $1.2{\cdot}10^{-1}$ | 33 (11) | 0.86 | 0.79 | - | [48] |
| p38 MAP kinase | MD sim. + IFP + PLS | $2.4{\cdot}10^{-6}$ - $2.8{\cdot}10^{-2}$ | 20 (6) | 0.72 | 0.56 | 0.06 | [49] |
| p38 MAP kinase | SMD | $2.0{\cdot}10^{-5}$ - $2.5{\cdot}10^{-1}$ | 8 | - | 0.88 | 4.50 | [56] |
| FAK/ PYK2 | tauRAMD | $5.4{\cdot}10^{-4}$ - $3.3{\cdot}10^{-1}$ | 24 | - | 0.91 | ~ $2.4{\cdot}10^{-3}$ - 0.12 | [61] |
| FAK | SMD | $2.6{\cdot}10^{-3}$ - 1.0 | 14 | - | 0.74 | NA[g] | [68] |

[a]Simulation time for one complex. Only production MD simulations are considered. System preparation and equilibration are excluded from this calculation. [b]Size of the data set. [c]Size of the test set, for methods which require training to predict kinetic rates (machine learning or chemometric methods). [d]$R^2$ value obtained using all complexes in the data set. [e]$R^2$ value obtained using p38 MAP kinase as a test set. [f]$R^2$ value obtained using FAK as a test set. [g]Not available.

## Concluding Remarks and Future Perspectives

The importance of binding kinetics for drug efficacy has led to the development of many computational methods for predicting kinetic rates in recent years. In this review, we summarized and compared some of the latest methods and applications based on two systems, trypsin-benzamidine and kinase-inhibitor complexes. Absolute kinetic rates could be estimated with low error with less than 1 microsecond of total simulation time by the use of methods such as M-WEM [20] and dcTMD [22], highlighting the advancements that can be achieved by second-generation methods. Such methods may soon compete, in terms of simulation time, with methods which predict relative kinetic rates and allow high-throughput simulations of large data sets, such as tauRAMD [61–63] and scaled MD [58,59]. While many methods are available to compute relative $k_{off}$ values, less methods to compute relative $k_{on}$ values are available, showing that enhancing sampling of ligand association can be more challenging. Computational methods can be used not only for prospective predictions of kinetic rates [69] but also to obtain mechanistic insights about selectivity [61] and drug resistance [44]. For machine learning, the availability of large and consistent data sets, obtained under similar experimental conditions, could help to generate better models and assist in understanding the limits of what can be achieved. Finally, the information provided here can lead to different ideas to improve predictions of kinetic rates with high errors, such as the use of scaling factors for methods which change the potential energy surface, careful parametrization of ligands, and extensive sampling of bound, transition and unbound states.

For the future, the limits of simulations for protein-ligand binding can be probed regarding system size and accuracy in the description of the potential energy (see Outstanding Questions box). A recent study [66], using bovine serum albumin as a crowder, suggested that the binding path of PP1 to Src kinase is modified in the presence of crowders. Previous studies [11,24,34] pointed out that the parameters to describe the potential energy of the system can in some cases negatively affect the accuracy of the predicted kinetic rates. We believe that the trend of advancement of computational methods for binding kinetics will gain more momentum and shape the future of drug discovery.

## Acknowledgments

Funding from DFG under Germany's Excellence Strategy – EXC 2008/1-390540038 – UniSysCat is gratefully acknowledged. We thank Dr. Daria Kokh for her helpful comments on the manuscript.

## Glossary

**CGMD**: Coarse-Grained Molecular Dynamics simulation, a method where atoms are grouped in beads in order to reduce the number of particles in the system.

**COMBINE:** Comparative Binding Energy analysis, a method where protein-ligand interaction energies are decomposed per residue and scaled using weighting factors obtained using partial least squares regression to reproduce experimental kinetic rates.

**dcTMD**: dissipation-corrected Targeted Molecular Dynamics.

**IFP**: interaction fingerprints.

**InMetaD**: infrequent metadynamics, a method where Gaussian functions are added to the potential energy of the system to flatten the free energy surface.

**LiGaMD**: Ligand Gaussian accelerated Molecular Dynamics, a method where the ligand nonbonded interaction potential energy is boosted to facilitate ligand dissociation, and the total potential energy is boosted to facilitate ligand association.

**M-WEM:** Markovian Weighted Ensemble Milestoning, a method which combines Markovian milestoning and weighted ensemble.

**MMVT**: Markovian Milestoning with Voronoi Tessellations.

**PLS**: partial least squares.

**REVO**: Resampling of Ensembles by Variation Optimization, a method based on WExplore.

**SEEKR**: Simulation Enabled Estimation of Kinetic Rates, a multiscale approach combining Brownian dynamics, molecular dynamics and milestoning.

**SMD**: steered molecular dynamics, a method where an external force is applied on an atom or a group of atoms to guide the system to the desired state. Here, force is applied on the ligand to facilitate ligand dissociation.

**SuMD**: Supervised Molecular Dynamics, a method in which a series of short simulations, e.g., 1 ns, are performed. After each short simulation, snapshots are selected according to a certain criterion and will be set as the starting points of the next short simulations. The criterion is chosen to guide the simulation towards unbinding events.

**tauRAMD**: tau-Random Acceleration Molecular Dynamics, a method where a force of constant magnitude and random orientation is applied to the ligand to facilitate dissociation.

**TMD**: Targeted Molecular Dynamics, a method where a constant velocity constraint is used to facilitate ligand dissociation. Experimental kinetic rates are correlated with the work obtained from simulations.

**UUMD**: Unaggregated Unbiased Molecular Dynamics, a method in which a high concentration of ligands (up to 60 mM) can be inserted into the simulation box without aggregation. Aggregation is prevented by the use of a small repulsive force between evenly distributed virtual interaction sites on the ligands.

**WE**: weighted ensemble, a method where a progress coordinate describing the event of interest is chosen and divided into bins. Every few iterations, trajectories are split or merged to keep a specific number of trajectories per bin.

**WEM**: weighted ensemble milestoning, a method in which weighted ensemble simulations are performed between the milestones.

**Figures**

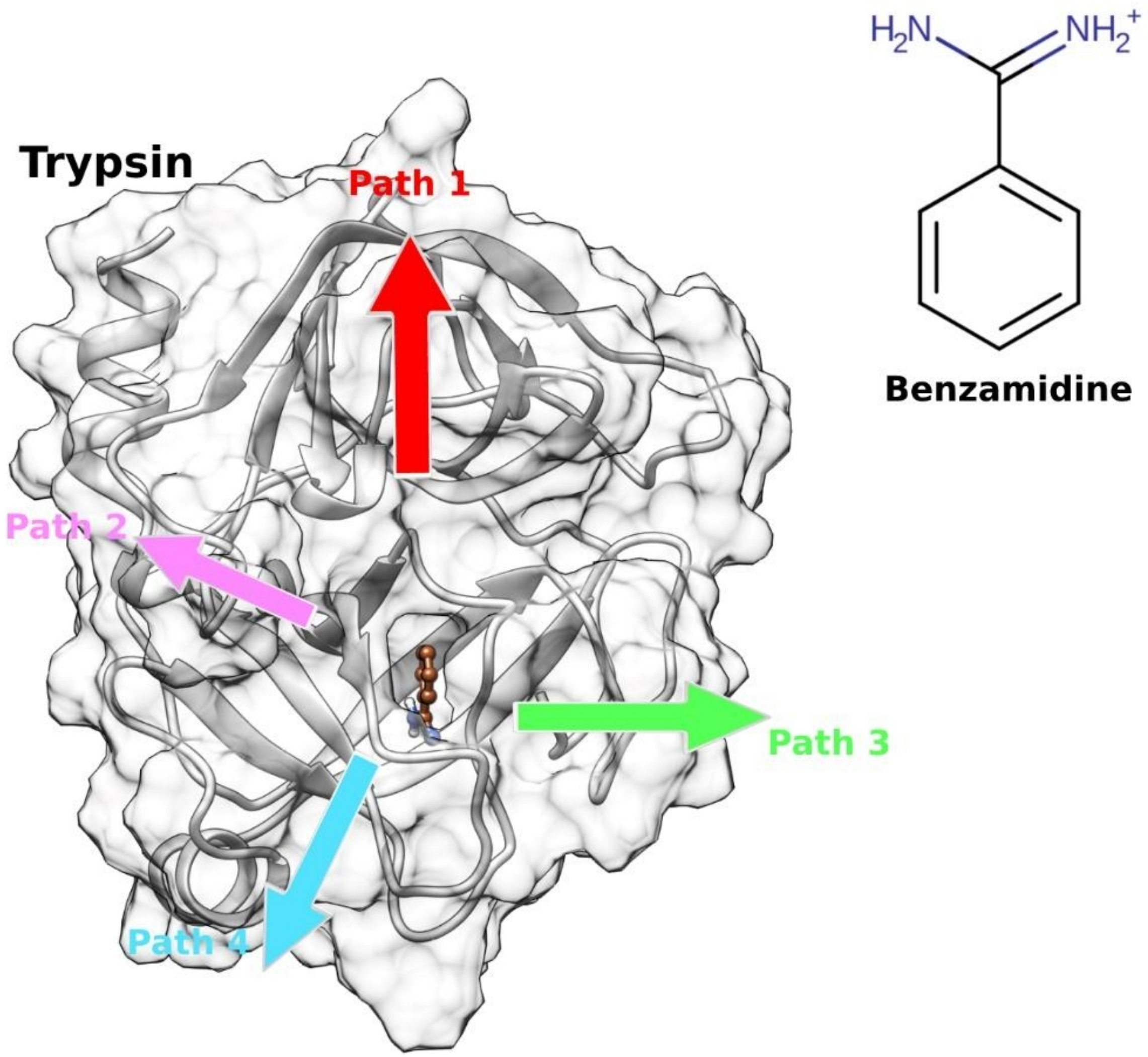

**Figure 1. Pathways for trypsin-benzamidine dissociation.** The crystallographic structure of trypsin in complex with benzamidine (PDB ID 1BTY [71]) and the 2D structure of benzamidine (protonation state at pH 7) are shown. The binding pocket is exposed to the solvent, facilitating ligand dissociation. A total of 4 unbinding paths, represented by the arrows, are characterized. Paths 1 and 2: REVO [23], WExplore [23], LiGaMD [21]; paths 3 and 4: REVO.

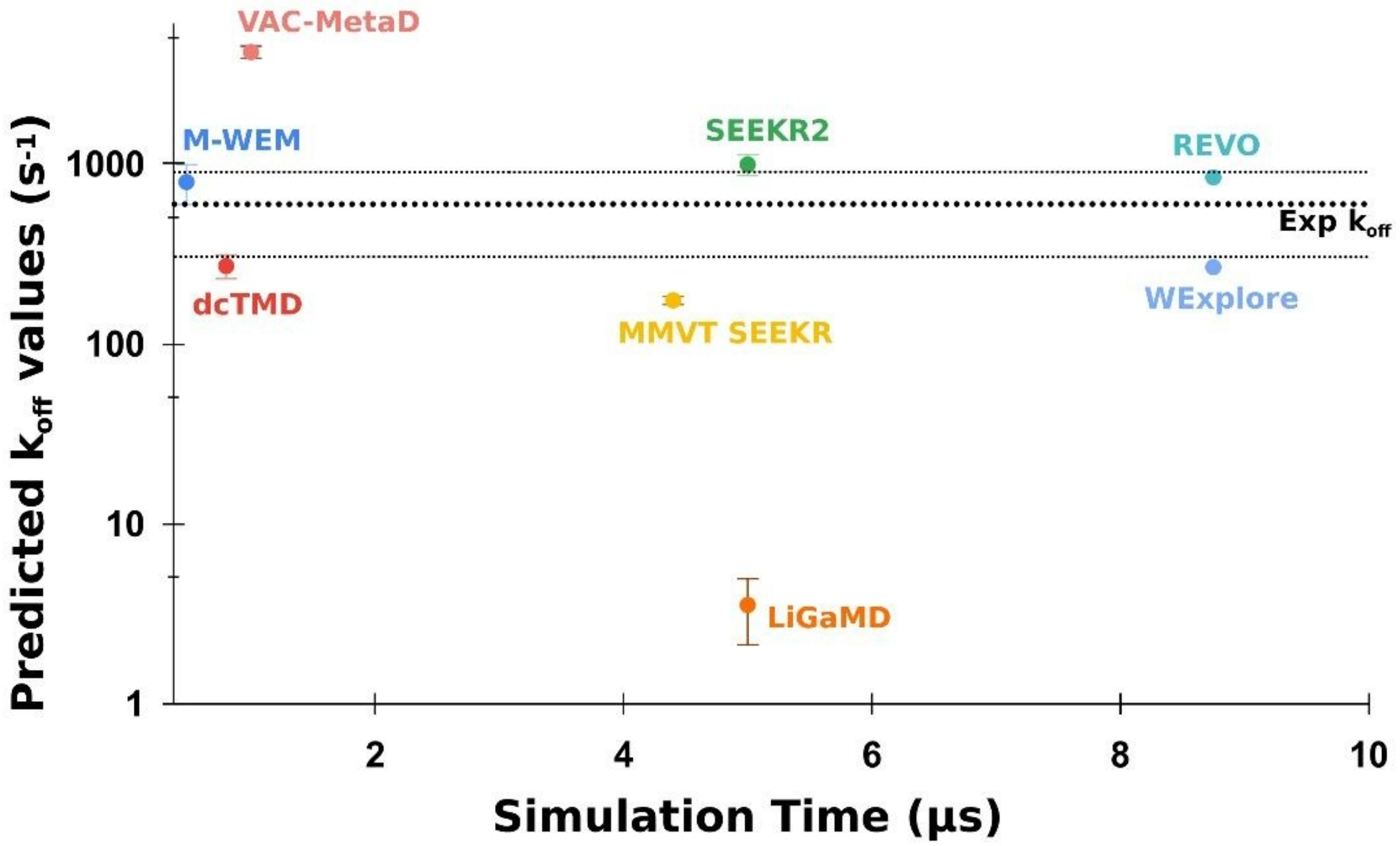

**Figure 2. An overview of the simulation time and predicted $k_{off}$ values for the trypsin-benzamidine complex, obtained from different methods, as shown in Table 1.** Each method is represented as a point. Only methods based on all-atom MD simulations are shown. The line with big dots represents the average experimental $k_{off}$ value (600 $s^{-1}$) [15] and the lines with small dots represent an error of 0.5 to 1.5-fold, or a range of ± 300 $s^{-1}$ around the average. The Y axis is in log scale.

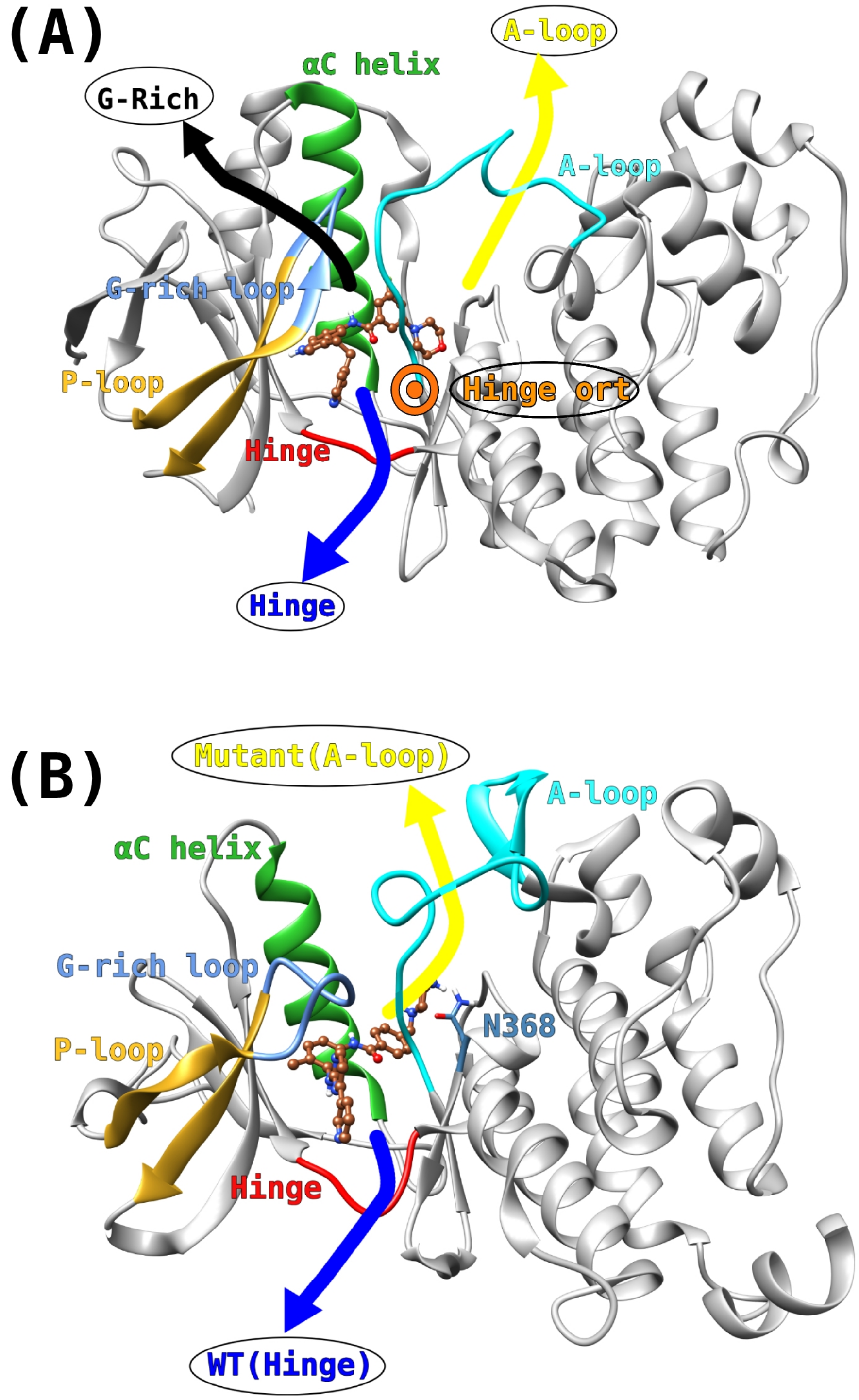

**Figure 3. Inhibitor unbinding pathways for p38 MAP kinase and Abl kinase**. (A) The crystallographic structure of p38 MAP kinase in complex with an inhibitor (PDB ID 1WBS [72]) is shown. A total of 4 unbinding paths, represented by the arrows, are characterized. ATP binding site (hinge) path: type I and type II inhibitors in SMD [56]

and local-scaled MD [53], type II inhibitors in metadynamics [57]; allosteric site (A-loop) path: type II inhibitors in metadynamics [57], SMD [56] and local-scaled MD [53]; path orthogonal to hinge (hinge ort): type I and type II inhibitors in local-scaled MD [53] (the circle indicates that the orthogonal hinge pathway is perpendicular to the readers' view); glycine-rich loop (G-rich) path: type II inhibitors in metadynamics [57]. (B) The crystallographic structure of Abl kinase in complex with Imatinib (PDB ID 1OPJ [73]) is shown. In a study which applied InMetaD to the Abl kinase, Imatinib displayed different unbinding pathways in the wild-type (hinge path) and in the drug-resistant mutant N368S (A-loop path) [44].